\documentclass[preprint,authoryear,12pt]{elsarticle}

\usepackage[unicode=true,
 bookmarks=true,bookmarksnumbered=false,bookmarksopen=false,
breaklinks=false,pdfborder={0 0 1},backref=false,colorlinks=false]
{hyperref}

\usepackage{amssymb}
\def\astrobj#1{#1}
\journal{New Astronomy}

\begin{document}

\begin{frontmatter}



\title{An Interacting \textsc{O+O} Supergiant Close Binary System: Cygnus OB2-5 (V729 Cyg)}


\author[ky1]{B. Ya{\c s}arsoy}
\author[ky1]{K.Yakut}
\address[ky1]{Department of Astronomy and Space Sciences, University of Ege, {\.I}zmir, Turkey}
%\address[ky2]{Institute of Astronomy, University of Cambridge, Madingley Road, Cambridge CB3 0HA, UK}

\begin{abstract}
The massive interacting close binary system \astrobj{V729 Cyg} (OIa+O/WN9), plausibly progenitor of a Wolf-Rayet system,
is studied using new observations gathered over 65 nights and earlier published data. Radial velocity and five colour light curves are analysed simultaneously.
Estimated physical parameters of the components are $M_1 = 36\pm 3 M_{\odot}$, $M_2 = 10 \pm 1 M_{\odot}$, $R_1 = 27\pm 1 R_{\odot}$, $R_2 = 15 \pm 0.6 R_{\odot}$,
$\log{(L_1/L_{\odot})} = 5.59\pm 0.06$, and $\log{(L_2/L_{\odot})}  = 4.65\pm 0.07$. We give only the formal 1$\sigma$ scatter, but we believe systematic errors in the luminosities,
of uncertain origin as discussed in the text, are likely to be much bigger. The distance of the \astrobj{Cygnus OB2} association is estimated as $967\pm 48$ pc by using our newly obtained parameters.
\end{abstract}


\end{frontmatter}

\section{Introduction}

One of the nearest ($d < 1.5$ kpc) OB associations, \astrobj{Cygnus OB2} contains
some of the most massive and luminous stars of our Galaxy. \astrobj{Cygnus OB2} most
probably belongs to \astrobj{Cygnus-X},  one of the richest massive star forming regions.
Earlier studies showed that the \astrobj{Cygnus OB2} region contains a number of
O, B, and Wolf-Rayet stars.
The total mass of the association is about 30,000 M$_{\odot}$ (Wright et al. 2012).
The age of the cluster was estimated to be in the range 1-10 Myr. Massey et al. (1995) gave
an age of 1-4 Myr and recently Wright et al. (2010) give its age as 3.5-5.3 Myr.
The \astrobj{Cygnus OB2} association is a good laboratory to study star formation and stellar evolution,
including the most advanced stages of the evolution of massive stars.
Hence, the association has been observed in $\gamma$-ray, X-ray, UV, optical, IR,
and radio wavelengths in numerous studies. Nevertheless the interstellar medium affects
the observations, especially in the UV and optical part of the electromagnetic spectrum (Rauw 2011).

The distance of the association has been calculated by various methods and these show
differences. Kiminki et al. (2007) calculated the visual extinction ($A_V$) as
5.4 mag and the distance modulus as 11.3 mag, by analyzing 146 OB systems.
Linder et al. (2009) have calculated the distance of the cluster as 925(25) pc from
a photometric analysis of the \astrobj{Cygnus OB2-5} system.
In the literature different results derived by various methods exist (see Rauw 2011).
These discrepancies possibly reflect the differences between the methods used as well
as the differences in the estimation of the visual extinction. In this study, we obtain
a new estimate of the distance, using our new photometric data of the eclipses

\astrobj{V729 Cyg} (\astrobj{Cygnus OB2-5}=BD+40$^\circ$4220, $P=6.6$ days) is a member of the association
and a massive and high-temperature contact binary (HTCB), and has been discussed in
many studies since it was discovered (e.g. Wilson 1948, Wilson \& Abt 1951, Hall 1974,
Bohannan \& Conti 1976, Leung \& Schneider 1978, Vreux 1985, Rauw et al. 1999,
Linder et al. 2009). The binary system consists of two supergiant components.
Recently, Kennedy et al. (2010) reported indirect evidence for the presence
of a third body in \astrobj{V729 Cyg} with an orbital period of 6.7 years.

The UBV light variations of the system were obtained by Hall (1974) and later these
light variations were used by Leung \& Schneider (1978) to model its light curve
using the Wilson-Devinney code (Wilson \& Devinney 1971, hereafter W-D code).
Leung \& Schneider (1978) then calculated the orbital and physical parameters of the system by using Bohannan \& Conti radial velocities.
They gave the masses of the components as $M_1 = 58.7\pm 9.1 M_{\odot}$
and $M_2 = 13.7\pm 6.3 M_{\odot}$.

Vreux (1985) reported a periodic change in the
H$_\alpha$ profile. Rauw et al. (1999) measured the radial velocities of the binary
components and studied the emission line profile variations caused by the stellar
wind interactions. They derived the mass functions for the components as
$M_1 \sin^3 i=24.6 M_{\odot}$ and $M_2 \sin^3 i=6.9 M_{\odot}$ and the mass loss
rate of the stellar winds as $\dot{M_1}=5.0\times10^{-6}M_{\odot}$/yr and
$\dot{M_2}=5.5\times10^{-6} M_{\odot}$/yr. They gave the spectral types of the
components as O6.5-7~Ia+Ofpe/WN9. Linder et al. (2009) investigated the light
variations of the system at 5057~{\AA} and 6051~{\AA}.
Assuming a bright spot Linder et al. modelled its LC with the Nightfall software.
The authors estimated the masses of components as $M_1 = 31.9\pm 3.2 M_{\odot}$
and $M_2 = 9.6\pm 1.1 M_{\odot}$, values with which we largely agree.

In this study newly obtained 5-colour (UBVRI) light curves of the system, published
UBV light curves of Hall (1974), and radial velocity curves from Rauw et al. (1999)
are analysed simultaneously and the orbital, physical parameters and the distance of the
association are presented. Following the observational information given in the second
section, light and radial velocity models are presented in the third section and
in the fourth section physical parameters of the binary system are calculated.

\section{New Observations}
The light variations of \astrobj{V729 Cyg} were observed in the Bessel U, B, V, R and I bands over
65 nights between August 2010 and October 2011. The observations were carried out at the
T\"UB\.ITAK National Observatory (TUG) with the 60cm telescope,
which was equipped with  an FLI CCD with $2048\times2048$ pixel.
Comparison stars selected from the literature were GSC 3161-01269 and GSC 03161-01384.
A total of 272, 417, 451, 281, and 280 data points were obtained in U, B, V, R and I bands, respectively.

The IRAF (DIGIPHOT/APPHOT) packages were used in data reduction.
The reduction and analysis of each frame are performed by subtracting the standard bias frames,
dark frames, and dividing by flat-field frames, followed by aperture photometry.
During the reduction, we have studied all the nights and each frame separately.
Standard deviations of the data are estimated as 0$^{\rm m}$.030, 0$^{\rm m}$.010, 0$^{\rm m}$.010,
0$^{\rm m}$.008, and 0$^{\rm m}$.009  for the $U$, $B$, $V$, $R$ and $I$ bands, respectively.

\begin{figure*}
\includegraphics[scale=0.8]{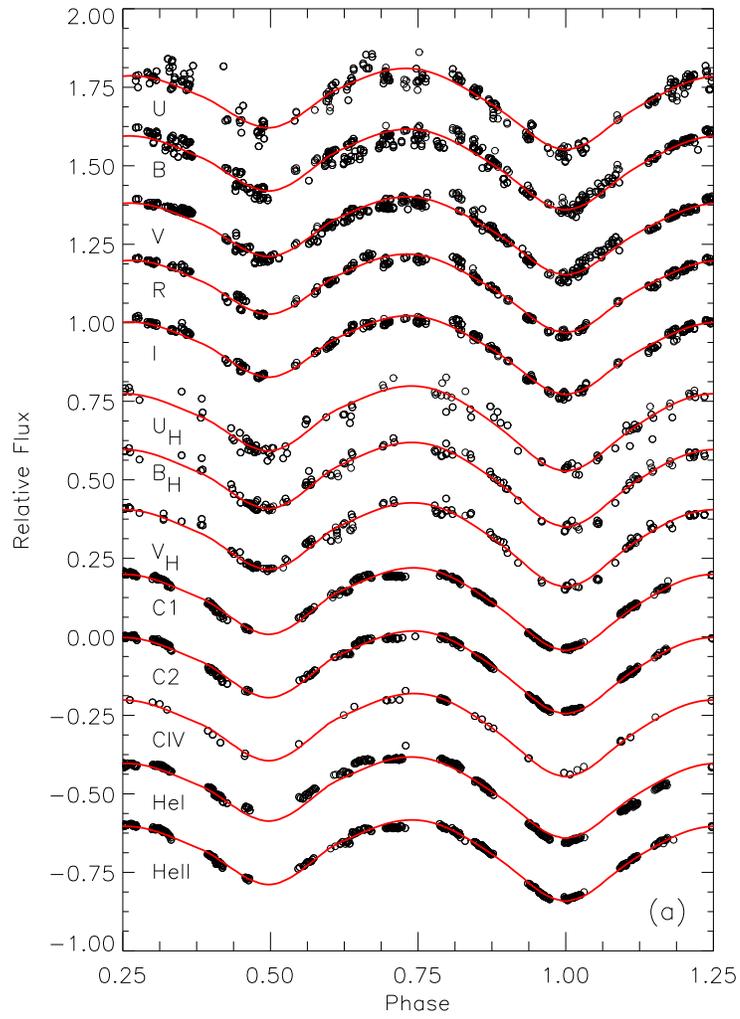}
\includegraphics[scale=0.8]{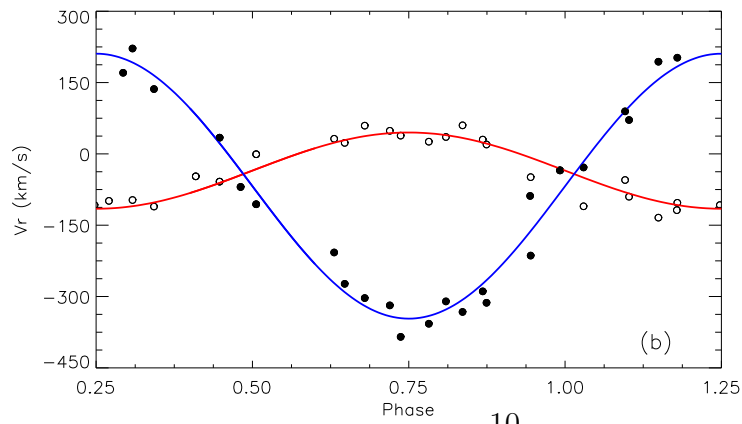}
\caption{(a) Observational (open circles) and computed (solid line) all available light (a) and radial velocity (b) curves of \astrobj{V729 Cyg}.}\label{fig:LC-WD}
\end{figure*}

All the new observations are given in Table~\ref{tab:obs}. The entire table can be found at
the CDS database. In Fig.~\ref{fig:LC-WD} we show the $UBVRI$ light curves of  \astrobj{V729 Cyg}.
The UBV light curves of Hall (1974) are also plotted in Fig.~\ref{fig:LC-WD} (namely as U$_H$, B$_H$, and V$_H$).
In Fig.~\ref{fig:LC-WD} we plotted phase {\it vs.} relative flux instead of magnitudes.
During data reduction, we used the linear ephemeris
\begin{equation}
\textrm{Min~I} = 24~40413.821(38) + 6.597887(28)\times E,
\label{Eq:1} \end{equation}
which is a reasonable approximation, over the last 40 years, to the quadratic ephemeris
obtained over 115 yrs in the next Section. Subsequently the code gave an improved
period of 6.597981, which is still in good enough agreement with our quadratic
ephemeris over the 40 yr time span.

\section{Eclipse timings and period study}
Detection of times of minima of long period eclipsing binary systems like \astrobj{V729 Cyg} have
relatively low probability. Hence the times of minima of \astrobj{V729 Cyg} obtained during the
last century are not as numerous as for other kinds of binaries. To search for the possible
causes of observed period variations one should study their O-C variations (O, Observed;
C, Calculated minima times).
This variations can bear information about the origin of the orbital period variation.
In this study, we collected all the minima times from the literature and list them in
Table~2, along  with the one obtained in this study. We have not included the minima
times of the system given by Kurochkin (1961) since they are given only to two digits
and are highly scattered. The data in Table~2 span over a 115 years, and so
covers an important amount of time in the context of high-mass
stellar evolution. The O-C variations show an upward parabolic trend. This indicates an
increase in orbital period, which can occur when mass transfer takes place from the
less massive star to the more massive one in a semidetached configuration; but a period
increase can also be expected if one or both components is losing mass to infinity through
stellar wind in a detached configuration. In addition if a third body orbits the binary
system there can be a long period high amplitude oscillation.  However, the possibility
of this is quite low since no such variation has been ever detected.

\begin{figure}
\includegraphics[width=90mm]{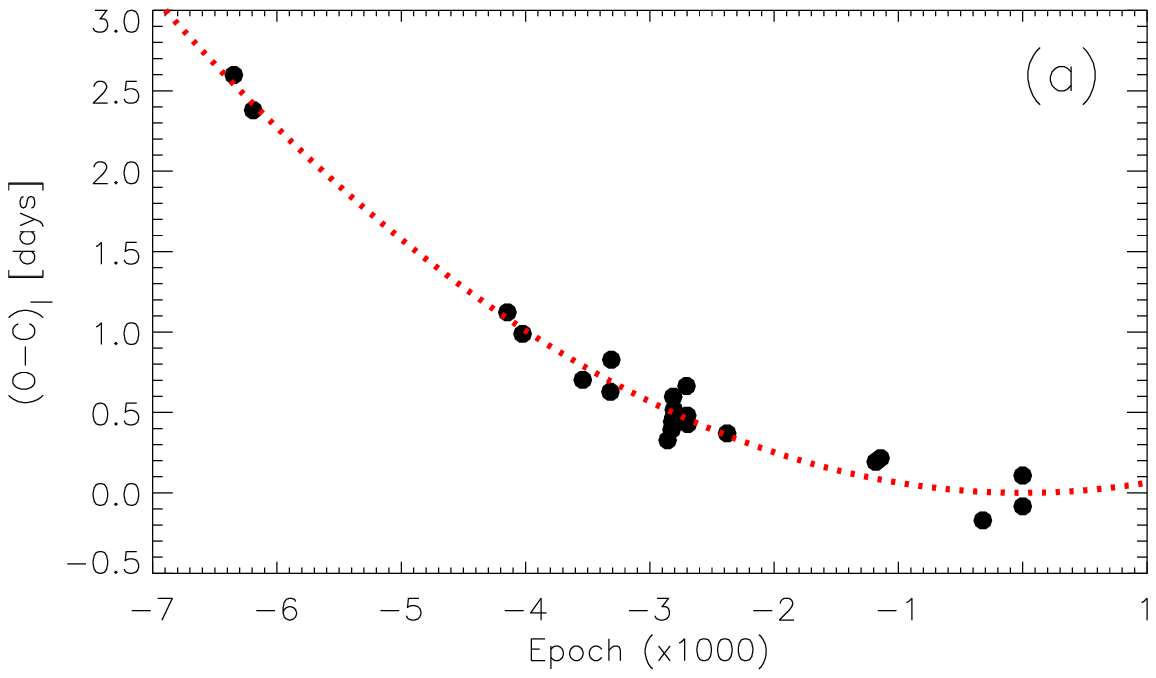} \\
\includegraphics[width=90mm]{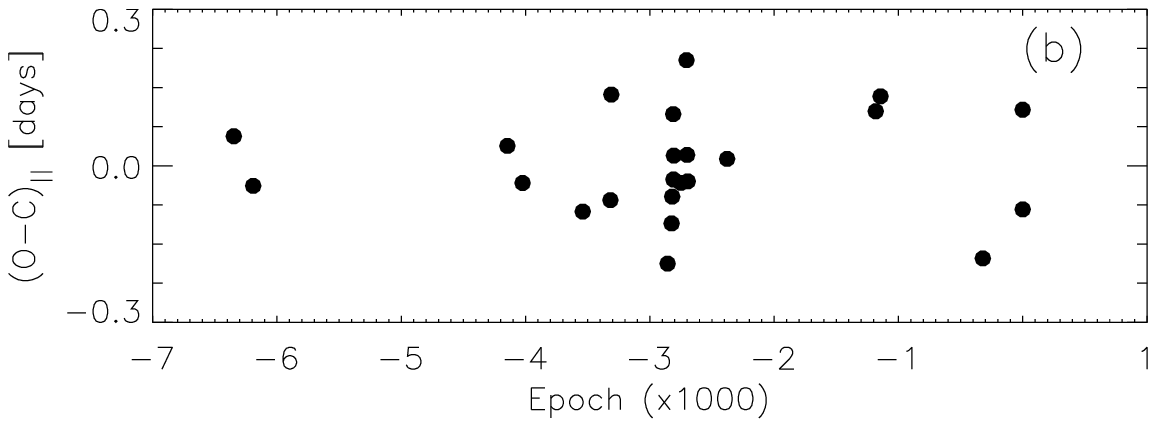}
\caption{ (a) The times of minimum light of \astrobj{V729 Cyg}. The dotted line
is the best-fit parabola. The total time interval is $\sim 115$ yr.
(b) The difference between the observations and the quadratic ephemeris.}
\label{Fig:V729Cyg:OC}
\end{figure}

A weighted least-squares method applied to the O-C timings  (Fig.~\ref{Fig:V729Cyg:OC}a)
gives a quadratic solution:

 \begin{equation}
\begin{array}{l}
\textrm{HJD} = {\rm HJD}_o +P_oE + {\frac{1}{2}}{\frac{dP}{dE}}E^2  \\
\\
= 24~56103.646(38) + 6.598074(12)\times E + 6.3(6)\times 10^{-8}\times E^2
 \label{Eq:V729cyg:1}
\end{array}
\end{equation}
The residuals are shown in Fig.~\ref{Fig:V729Cyg:OC}b. This result indicates an
orbital period variation on a timescale $P/\dot{P}\sim 9.4\times 10^{5}$. We do not
detect the third-body period of $6.7$ yrs suggested by Kennedy et al (2010).

\section{Light curve modeling}

We have reduced the observed magnitude to relative flux before light variation analysis.
We did a similar process to the Hall (1974) UBV data. Later we determined the weights
by taking in to consideration the observational errors.
According to this we determined the weights for U, B, V, R, and I filters
respectively as  3.3, 10, 10, 12.5, and 11.1. We used the Phoebe (Pr\~{s}a \&
Zwitter 2005) program based on the W-D code. In the analysis the radial velocity data
from Rauw et al. (1999), as well as the Hall (1974) UBV observations
and the UBVRI observations obtained in this study were used.

During the analysis the temperature of the primary component, the limb darkening
(van Hamme 1993), albedos (Rucinski 1969) and gravity darkening (von Zeipel, 1924)
are regarded as fixed parameters. The temperature of the primary component was
assumed as either 36000~K from the spectral study of Rauw et al. (1999), or 32000~K.
The latter value is appropriate for O9I (Cox 2000).
The time of minimum light $T_0$, orbital period $P$, orbital inclination $i$,
mass ratio $q$, temperature of the secondary component $T_2$, surface potentials
$\Omega_1 = \Omega_2$, luminosities $L_1$(U), $L_1$(B), $L_1$(V), $L_1$(R), $L_1$(I),
$L_1$(U1), $L_1$(B1), $L_1$(V1) are free parameters to be solved for. The results are
summarized in Table~\ref{tab:lc-results}. The analysis indicates
that the light contribution of the primary component is 86\% in U, 85\% in B, V and R,
and 84\% in the I-light curve, for our 32000K solution. In Fig.~\ref{fig:LC-WD} the
light curves computed with the resulting parameters are shown by solid lines. Modeling
of the system indicates Roche-lobe overfilling components. The filling factor, $f = 0.22$
or 0.17, is given by
$(\Omega _{\textrm{in}} -\Omega)/ (\Omega _{\textrm{in}} -\Omega _{\textrm{out}} )$,
and varies from zero to unity from the inner to the outer critical surface.
This solution indicates that the system has a moderate degree of contact.

\section{Astrophysical parameters of the system}
\def\vg{V_{\gamma}}
The speed of the center of mass,
$\vg$, averages as  $-55$. That the two components are best represented by
somewhat different $\vg$'s probably represents the fact that winds from each component
have different speeds, and so distort and shift the lines differently. Although
the colors such as B-V vary rather little round the orbit, as can be seen in Fig. 1a,
the temperatures are required to differ because of the different geometries of the
two eclipses.

The physical parameters of \astrobj{V729 Cyg} listed in Table~\ref{tab:PhyPar} are determined
from the parameters of Table~\ref{tab:lc-results}. The temperature of the Sun was
taken as 5777~K and its bolometric magnitude as 4.732 mag.

In order to obtain the absolute visual magnitude the bolometric corrections (BC)
are estimated from Martins et al. (2005).
The total magnitude of the system is taken as V=9.21 mag (Hall 1974), and the
individual magnitudes as 9.41 and 11.15.
The $E_{B-V}$ value (1.99 $\pm$ 0.03) of the system was obtained from
Friedman et al. (2011).  Using the apparent magnitude of the system and the values
quoted in Table ~\ref{tab:PhyPar}, we infer the distance of the system to be
967 $\pm$ 48 pc. Distance estimation from the binary parameters gives a result
that is consistent with the other distance estimation techniques (see Section 1).

\section{Results and Conclusion}

In this study we investigated the massive interacting close binary
system \astrobj{V729 Cyg} by using our new and earlier published observations.
Radial velocity and light curves of the system were solved simultaneously,
and the orbital and physical parameters obtained are shown in
Table~\ref{tab:lc-results} and Table~\ref{tab:PhyPar}.
Our analysis shows that the system has a moderate contact configuration.

In their studies on X-ray and radio emission, Linder et al. (2009) and Kennedy
et al. (2010) suggested that there is a third body with a period of 6.7 years
orbiting the system \astrobj{V729 Cyg} from long-term VLA observations. Kennedy et al. (2010)
estimated the mass function related to the third body as 3.2  $\rm{M_{\odot}}$.
By using this mass function value and the physical parameters given in
Table ~\ref{tab:PhyPar} the mass of the third body for
$i_3$=90$^o$, 70$^o$, 50$^o$, 30$^o$ is obtained respectively as 24
$\rm{M_{\odot}}$, 26 $\rm{M_{\odot}}$, 35 $\rm{M_{\odot}}$, 68 $\rm{M_{\odot}}$.
We would like to emphasize that the uncertainty in these mass values is high
because of the uncertainty in the mass function.

Single massive stars above an initial mass of $8M_{\odot}$ are known to evolve towards an explosive
supernova explosion (Smartt et al. 2009). Leaving a neutron star or black hole remnant.
However if a massive stars is a member of binary like \astrobj{V729 Cyg} its evolutionary path towards the final
supernovae will become significantly different. The nature of the final stellar remnant may also be altered
so that while a neutron star might be expected mass transfer may lead to black hole formation.
Observations of systems such as \astrobj{V729 Cyg} will allow us to refined out binary evolutionary
models to increase our understanding of the possible variety of stellar lifecycles (see Eggleton 2010, Eldridge \& Stanway 2009, Yakut \& Eggleton 2005, Pols et al. 1995)

\section*{Acknowledgments}
The authors thank Peter Eggleton, Chris Tout and Belinda Kalomeni for their valuable comments
and suggestions to improve the quality of the paper. We are grateful
to an anonymous referee for comments and suggestions which helped us to improve the paper.
This study was supported by the Turkish Scientific and Research Council (T\"UB\.ITAK 111T270 and 113F097) and
the T\"UB\.ITAK National Observatory (T60-Pr68).
KY thanks the Institute of Astronomy, University of Cambridge for its support during his visit to IoA.
The current study is a part of PhD thesis of B. Ya{\c s}arsoy.

\begin{table}
\caption{UBVRI measurements of \astrobj{V729 Cyg} (Fig.~\ref{fig:LC-WD}). The phases were
calculated using Eq. (1). 1, 2, 3, 4, and 5 denote U, B, V, R, and I filters (F),
respectively. All data is published in its entirety at The Strasbourg astronomical Data Center (CDS).
A part of table is shown here for guidance regarding its form and content.} \label{tab:obs}
\vskip 0.1truein
\begin{tabular}{llll}
\hline
\hline
HJD	        &	Phase	&	$\Delta$m	&	F \\	
24~00000+   &	 	    &	mag       	&	 \\	
\hline
53301.3054	&	0.5115	&	1.2930	&	1	\\
53301.3067	&	0.5151	&	1.2840	&	1	\\
53301.3080	&	0.5187	&	1.2610	&	1	\\
53301.3093	&	0.5224	&	1.2500	&	1	\\
53301.3106	&	0.5260	&	1.2300	&	1	\\
53301.3119	&	0.5296	&	1.2050	&	1	\\
53301.3133	&	0.5333	&	1.1840	&	1	\\
53301.3146	&	0.5369	&	1.1690	&	1	\\
53301.3159	&	0.5405	&	1.1560	&	1	\\
\hline
\end{tabular}
\end{table}

\begin{table}
\caption{Times of minimum light for \astrobj{V729 Cyg}.}\label{tab:V729Cyg:mintimes}
\vskip 0.1truein
\begin{tabular}{llll}
\hline
HJD Min    &   Ref &   HJD Min    &   Ref\\
\hline
14228.269 & 1 & 37563.530 & 4  \\
15257.350 & 1 & 37583.369 & 3  \\
28749.154 & 1 & 37959.387 & 3  \\
29553.985 & 1 & 38256.520 & 3  \\
32747.167 & 2 & 38289.327 & 3  \\
34218.463 & 2 & 38322.265 & 3  \\
34264.849 & 1 & 40413.796 & 2  \\
37253.276 & 3 & 48301.616 & 5  \\
37464.480 & 4 & 48555.665 & 5  \\
37497.520 & 4 & 53985.493 & 6  \\
37550.460 & 4 & 56100.454(4) & 7  \\
          &   & 56103.562(3) & 7  \\
\hline
\end{tabular}
\vskip 0.1truein
{References for Table~\ref{tab:V729Cyg:mintimes}.
1-  Sazonov (1961)
2-  Hall (1974)
3-  Haeussler (1964)
4-  Romano (1969)
5-  ESA, Hipparcos
6-  Heubscher \& Walter (2007)
7- present study.}
\end{table}

\begin{table*}
\begin{center}
\scriptsize
\caption{The photometric and spectroscopic elements of \astrobj{V729 Cyg} with their formal
1$\sigma$ errors. See text for details.}\footnotesize \label{tab:lc-results}
\vskip 0.1truein
\begin{tabular}{llll}
\hline
Parameter                                   &  T$_1=28000$~K    &  T$_1=32000$~K     &  T$_1=36000$~K     	 \\
\hline
T$_{\rm o}$                                 & 24~40413.799(5)   & 24~40413.798(5)    &  24~40413.799(5)     \\
P  (day)                                    & 6.597983(3)       & 6.597982(2)        &  6.597982(2)         \\
Spectroscopic parameters:&&\\
$K_1$                                       & 82.0(2.3)         & 79.9(2.2)          & 88.0(2.3)            \\
V$_{\gamma 1}$                              & -33.3(3.3)        & -33.3(3.3)          & -35.10(3.3)          \\
$K_2$                                       & 282.9(8.3)        & 285.2(8.6)         & 285.7(8.6)             \\
V$_{\gamma 2}$                              & -68.0(6.7)        & -68.0(6.7)         & -67.8(6.7)         	  \\
Geometric parameters:&&&                                                                                       \\
i ${({^\circ})}$                            & 63.8(1)           & 63.9(1)            & 64.1(1)     				\\
a                                           & 53(2)             & 53(2)              & 53(2)                    \\
q                               			& 0.290(12)         & 0.290(12)          &               		 	 \\
$\Omega _{1,2}$                             & 2.406(4)          & 2.408(5)           & 2.414(7)         		\\
Filling factor (\%)                         & 21(2)             & 20(2)              & 17(3)                	\\
Fractional radii of the primary component   &&&\\&&&\\
r$_{1~\textrm{pole}}$                       &0.4665             &0.4661              &0.4647           	        \\
r$_{1~\textrm{side}}$                       &0.5042             &0.5036              &0.5017            	   \\
r$_{1~\textrm{back}}$                       &0.5318             &0.5311              &0.5286                    \\
$\bar{r}_{1}$ (volume radius)               &0.5001(4)          &0.4996(5)           &0.4977(6)              \\
Fractional radii of the secondary component &&& \\
r$_{2~\textrm{pole}}$                       &0.2672             &0.2668              &0.2653                \\
r$_{2~\textrm{side}}$                       &0.2794             &0.2789              &0.2771                 \\
r$_{2~\textrm{back}}$                       &0.3193             &0.3184              &0.3151                 \\
$\bar{r}_{2}$                               &0.2878(4)          &0.2872(6)           &0.2851(6)               \\
Radiative parameters:&&&\\
%T$_1$$^*$ (K)                              & 28000$^*$         & 32000$^*$          & 36000$^*$             \\
T$_2$ (K)                                   & 21540(375)        & 24085(420)         & 26620(480)            \\
Albedo$^*$ ($A_1=A_2$)                      & 1.0               & 1.0                &  1.0                   \\
Gravity brightening$^*$ ($g_1=g_2$)         & 1.0               & 1.0                &  1.0                    \\
\hline $^*$~Fixed
\end{tabular}
\end{center}
\end{table*}

\begin{table*}
\begin{center}
\caption{Absolute parameters of \astrobj{V729 Cyg}. The standard 1$\sigma$ errors in the last digit are
given in parentheses.}\label{tab:PhyPar}
\vskip 0.1truein
\begin{tabular}{llll}
\hline
Parameter                              &Unit             & Pri. (OIa)  	& Sec. (O/WN9)\\
\hline
Mass (M)                               &$\rm{M_{\odot}}$ & 31.6(2.9)   	& 8.8(3)      \\
Radius (R)                             &$\rm{R_{\odot}}$ & 25.6(1.1)   	& 14.5(1.0)   \\
Temperature (T$_{\rm eff}$)            &$\rm{K}$         & 28000       	& 21263(370)  \\
Luminosity ($\log L$)                  &$\rm{L_{\odot}}$ & 5.558(65)   	& 4.587(96)   \\
Surface gravity      ($\log g$)        & CGS             & 3.12(2)     	& 3.06(2)     \\
Absolute bol. magnitude (M$_b$)        &mag              & -9.16       	& -6.74       \\
Absolute vis. magnitude (M$_V$)        &mag              & -6.50       	& -4.89       \\
Separation between stars ($a$)         &$\rm{R_{\odot}}$ &  ~~~~~~~~~50.8(1.8)   	&                \\
Distance ($d$)                         & pc              & ~~~~~~~~~932(20)     	&             \\
\hline
\end{tabular}
\end{center}
\end{table*}


\begin{thebibliography}{}
\bibitem[\protect\citeauthoryear{Bohannan \& Conti}{1976}]{1976ApJ...204..797B} Bohannan B., Conti P.~S., 1976, ApJ, 204, 797
\bibitem[\protect\citeauthoryear{Cox}{2000}]{2000asqu.book.....C} Cox A.~N., 2000, Allen's astrophysical quantities, 4th ed. Publisher: New York: AIP Press; Springer
\bibitem[\protect\citeauthoryear{Eggleton}{2010}]{2010NewAR..54...45E} Eggleton P.~P., 2010, NewAR, 54, 45
\bibitem[\protect\citeauthoryear{Eldridge \& Stanway}{2009}]{2009MNRAS.400.1019E} Eldridge J.~J., Stanway E.~R., 2009, MNRAS, 400, 1019
\bibitem[\protect\citeauthoryear{ESA}{1997}]{1997yCat.1239....0E} ESA, 1997, yCat, 1239, 0
\bibitem[\protect\citeauthoryear{Friedman et al.}{2011}]{2011ApJ...727...33F} Friedman S.~D., et al., 2011, ApJ, 727, 33
\bibitem[\protect\citeauthoryear{Hall}{1974}]{1974AcA....24...69H} Hall D.~S., 1974, AcA, 24, 69
\bibitem[Haeussler99]{K} Haeussler, K., Lich, HABZ 23
\bibitem[Hubscher \& Walter(2007)]{2007IBVS.5761....1H} Hubscher, J., \& Walter, F.\ 2007, Information Bulletin on Variable Stars, 5761, 1
\bibitem[\protect\citeauthoryear{Kennedy et al.}{2010}]{2010ApJ...709..632K} Kennedy M., Dougherty S.~M., Fink A., Williams P.~M., 2010, ApJ, 709, 632
\bibitem[\protect\citeauthoryear{Kiminki et al.}{2007}]{2007ApJ...664.1102K} Kiminki D.~C., et al., 2007, ApJ, 664, 1102
\bibitem[\protect\citeauthoryear{Leung \& Schneider}{1978}]{1978ApJ...224..565L} Leung K.-C., Schneider D.~P., 1978, ApJ, 224, 565
\bibitem[\protect\citeauthoryear{Linder et al.}{2009}]{2009A&A...495..231L} Linder N., Rauw G., Manfroid J., Damerdji Y., De Becker M., Eenens P., Royer P., Vreux J.-M., 2009, A\&A, 495, 231
\bibitem[Martins et al.(2005)]{2005A&A...436.1049M} Martins, F., Schaerer, D., \& Hillier, D.~J.\ 2005, A\&A, 436, 1049
\bibitem[\protect\citeauthoryear{Massey, Johnson, \& Degioia-Eastwood}{1995}]{1995ApJ...454..151M} Massey P., Johnson K.~E., Degioia-Eastwood K., 1995, ApJ, 454, 151
\bibitem[\protect\citeauthoryear{Pols et al.}{1995}]{1995MNRAS.274..964P} Pols O.~R., Tout C.~A., Eggleton P.~P., Han Z., 1995, MNRAS, 274, 964
\bibitem[\protect\citeauthoryear{Pr{\v{s}}a \& Zwitter}{2005}]{prs05} Pr{\v{s}}a, A., Zwitter, T., 2005, ApJ, 628, 426
\bibitem[\protect\citeauthoryear{Rauw}{2011}]{2011A&A...536A..31R} Rauw G., 2011, A\&A, 536, A31
\bibitem[\protect\citeauthoryear{Rauw, Vreux, \& Bohannan}{1999}]{1999ApJ...517..416R} Rauw G., Vreux J.-M., Bohannan B., 1999, ApJ, 517, 416
\bibitem[\protect\citeauthoryear{Romano}{1969}]{1969MmSAI..40..375R} Romano G., 1969, MmSAI, 40, 375
\bibitem[\protect\citeauthoryear{Ruci{\'n}ski}{1969}]{1969AcA....19..245R} Ruci{\'n}ski S.~M., 1969, AcA, 19, 245
\bibitem[\protect\citeauthoryear{Sazonov}{1961}]{1961PZ.....13..445S} Sazonov V., 1961, PZ, 13, 445
\bibitem[\protect\citeauthoryear{van Hamme}{1993}]{1993AJ....106.2096V} van Hamme W., 1993, AJ, 106, 2096
\bibitem[\protect\citeauthoryear{von Zeipel}{1924}]{ze}von Zeipel, H., 1924, MNRAS, 84, 665
\bibitem[\protect\citeauthoryear{Vreux}{1985}]{1985A&A...143..209V} Vreux J.~M., 1985, A\&A, 143, 209
\bibitem[\protect\citeauthoryear{Wilson \& Abt}{1951}]{1951ApJ...114..477W} Wilson O.~C., Abt A., 1951, ApJ, 114, 477
\bibitem[\protect\citeauthoryear{Wilson}{1948}]{1948PASP...60..385W} Wilson O.~C., 1948, PASP, 60, 385
\bibitem[\protect\citeauthoryear{Wilson \& Devinney}{1971}]{wil71} Wilson, R.E., Devinney, E.J., 1971, ApJ, 166, 605
\bibitem[\protect\citeauthoryear{Wright et al.}{2010}]{2010ApJ...713..871W} Wright N.~J., Drake J.~J., Drew J.~E., Vink J.~S., 2010, ApJ, 713, 871
\bibitem[\protect\citeauthoryear{Wright et al.}{2012}]{2012scel.book..179W} Wright N.~J., Drake J.~J., Drew J.~E., Vink J.~S., 2012, scel.book, 179
\bibitem[\protect\citeauthoryear{Yakut \& Eggleton}{2005}]{YE05} Yakut K., Eggleton P.~P., 2005, ApJ, 629, 1055
\end{thebibliography}
\end{document}